\documentclass[a4paper]{article}

\usepackage{INTERSPEECH2020}
\usepackage{rotating}
\usepackage{tikz}
\usepackage{multicol}

\title{Identify Speakers in Cocktail Parties with End-to-End Attention}
\name{Junzhe Zhu, Mark Hasegawa-Johnson, and Leda Sar\i}
\address{ECE Department, University of Illinois, USA}
\email{\{junzhez2,jhasegaw,lsari2\}@illinois.edu}

\begin{document}

\maketitle
\begin{abstract}
  In scenarios where multiple speakers talk at the same time, it is important to be able to identify the talkers accurately. This paper presents an end-to-end system that integrates speech source extraction and speaker identification, and proposes a new way to jointly optimize these two parts by max-pooling the speaker predictions along the channel dimension. Residual attention permits us to learn spectrogram masks that are optimized for the purpose of speaker identification, while residual forward connections permit dilated convolution with a sufficiently large context window to guarantee correct streaming across syllable boundaries.  End-to-end training results in a system that recognizes one speaker in a two-speaker broadcast speech mixture with 99.9\% accuracy and both speakers with 93.9\% accuracy, and that recognizes all speakers in three-speaker scenarios with 81.2\% accuracy.\footnote{Code: https://github.com/JunzheJosephZhu/Identify-Speakers-in-Cocktail-Parties-with-E2E-Attention.git}
\end{abstract}
\noindent\textbf{Index Terms}: speaker recognition, source separation, cocktail party effect

\section{Introduction}
\label{sec:introduction}

The ``Cocktail Party'' is the problem of recognizing the identity of multiple speakers who are talking at the same time; in order to distinguish it from other types of simultaneous-speech problems, this problem is also called co-channel speaker identification~\cite{Morgan1995}.  Different methods have been proposed for scenarios where there are multiple microphone channels, or when there are visual inputs~\cite{Ephrat_2018}, but identification of multiple speakers from a monaural signal is challenging, because the same band now contains sounds from different sources~\cite{Allen94}.  Full-band speaker identification technologies, like Gaussian mixture models (GMM)~\cite{Reynolds95} and i-vectors~\cite{5545402}, fail.  Early approaches to co-channel speaker identification simply tried to identify temporal segments sufficiently dominated by one speaker, to permit single-speaker identification~\cite{Krishnamachari2000}, but since single-channel source separation became possible~\cite{Brown94}, its methods were applied to the problem of co-channel speaker ID~\cite{Shao2003}.  Automatic speaker identification systems based on GMMs were applied using the fusion of speaker-dependent models~\cite{Mowlaee2010}, using the inter-GMM Kullback-Leibler divergence~\cite{Saeidi2010}, using vector quantization~\cite{Li2010}, and by integrating the GMMs in a graphical model~\cite{Hershey2010}.

In a wide variety of applications recently, it has been demonstrated that end-to-end training of a deep neural network can result in lower error rates than a system built up from separately optimized components~\cite{Subramanian2020,Sari2020}.  End-to-end co-channel speaker identification is also possible: the speaker identification loss can be back-propagated to train the weights of the source separation algorithm.  Multi-speaker recognition is, obviously, a type of multi-label classification problem~\cite{Tsoumakas2007}, and it is possible to train a co-channel speaker ID system using multi-label classification methods~\cite{Khademian2018}.  Improved accuracy is possible, however, if one can force the network to output the correct number of speakers, rather than allowing the network to choose the number of speakers.  Doing so requires solving the streaming problem (the problem of correctly assigning each decoded feature vector to a feature stream) and the permutation problem (the problem of computing gradients based on the correct assignment of feature streams to ground truth speaker identities).  The streaming problem can be solved using stream-assignment methods from computational auditory scene analysis~\cite{Liu2018}, or using dilated convolution~\cite{Wang2018}.  The permutation problem can be solved using a dominant-speaker model~\cite{Jansky2019}, but is usually solved using some variant of permutation-invariant training (PIT)~\cite{Yu2017,Liu2018}, in which feature streams are assigned to ground truth reference signals in whatever order minimizes the reconstruction error.

This paper develops a new algorithm for co-channel speaker identification, based on methods of residual attention, dilated convolution with residual connections, Siamese networks for speaker identification, and a novel max-pooling error metric.
Residual attention computes a spectrogram transformation and spectrogram mask that are simultaneously optimized~\cite{Huang2015}.
Residual connections permit us to implement a much deeper dilated convolution than the one proposed in~\cite{Wang2018} (18 layers compared to their 3 layers), thereby permitting a more robust streaming of the segregated features.
Siamese networks are used to pre-train the speaker recognition network~\cite{Chen2011,Sari2019}.
The entire end-to-end network is then trained for optimum co-channel speaker identification using
a max-pooling error criterion that ignores permutation, in a manner similar to PIT~\cite{Yu2017}, but without PIT's computational expense.

\section{Source extractor model}
\label{sec:extractor}

\begin{figure*}[t]
    \center
     \includegraphics[width=\textwidth]{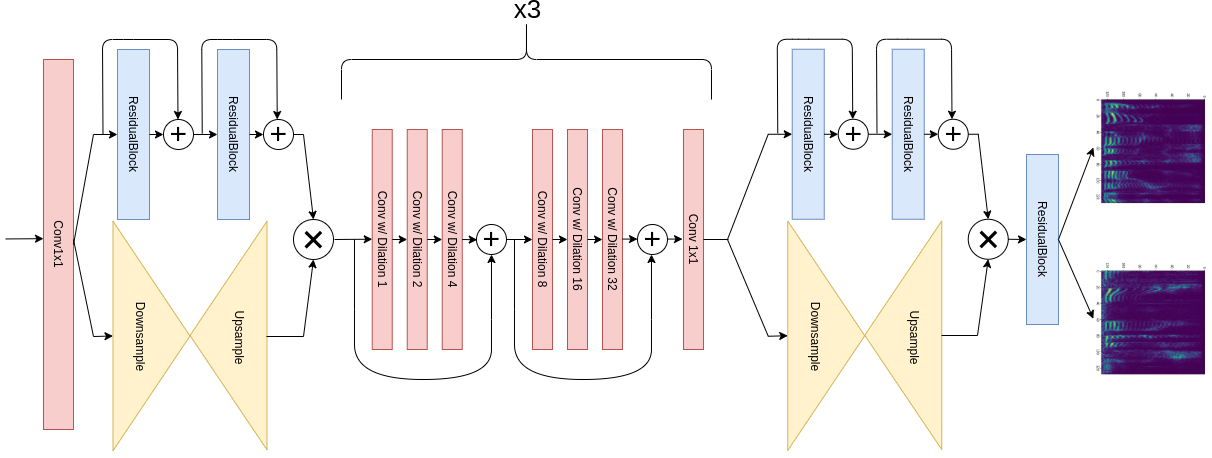}
    \caption{Extractor module for 2-speaker model. The residual blocks are the 'trunk' branch, while the hourglass architecture is the 'mask' branch. All convolutions are 2D.}
    \label{fig:1}
\end{figure*}

The first stage of our pipeline is an extractor that uses residual attention and dilated convolution to extract and stream individual source spectrograms from the mixed speech spectrogram.

The scaling of speech spectrograms is problematic. The magnitude STFT requires different network sensitivity at different frequencies: amplitudes of low-frequency pitch lines are dozens of times higher than those of high-frequency pitch lines. The decibel scale is unsatisfactory: logarithms of small numbers are large negative numbers.  Standard spectrogram visualizations (e.g.~\cite{Potter66}) truncate the spectrogram 60dB below its maximum, but the statistics of a truncated input are difficult for a neural network to learn.  A useful balance is obtained using $X=\log(1+S)$ as the network input, where $S=|STFT|$ is the magnitude of the short-time Fourier transform for mixed and separated speech signals.

The task of separating a mixed spectrogram boils down to assigning pitch lines, frication frames, and aspiration frames correctly to each source, then refining the time-frequency bins that are under interference. Therefore, we hypothesize that in order to separate the sources as closely as possible, the neural network needs to first have some higher-level knowledge about each source, in order to correctly identify features from different sources. For this purpose we employ residual attention~\cite{Wang2017} as the main backbone of the model; its hourglass architecture collects contextual information and generates masks that are able to suppress fine-grained regions of the spectrogram. 

The extractor model begins and ends with residual attention networks (the dual-path sections at left and right sides of Fig.~\ref{fig:1}).  Residual attention computes two different types of nonlinear tensor transformations: a trunk branch, $T(X)$, and a set of attention masks, $M(X)$.  The trunk branch ($T(X)$) is a sequence of convolutional layers with residual connections, shown in the top path of Figure~\ref{fig:1}.  The masks ($M(X)$) are computed in a series of downsample-then-upsample operations, arranged in a U-Net~\cite{Ronneberger2015} architecture, shown in the bottom path of Figure~\ref{fig:1}.  The U-Net architecture learns global structure (e.g., harmonic structure, and the alternation of voiced and unvoiced speech frames) with which to estimate $M(X)$.  The residual attention layers compute $(1+M(X))\odot T(X)$, an initial estimate of the separated speech sources, where the plus term represents a residual connection for better gradient propagation.    Several residual attention architectures were tested; the best-performing system uses residual attention blocks with 128 channels and 8x downsample-upsample ratio in the mask branch.


The residual attention block ensures consistency of the harmonic structure in each continuously-voiced interval, but does not ensure consistency of speaker identity between voiced regions.  A convolution kernel composed mostly of zeros, with nonzero values only once every $M$ samples, is called a dilated convolution~\cite{Vaidyanathan93}.  It has been demonstrated that dilated convolutional layers, with dilation factors that increase exponentially from layer to layer, can accumulate long-range context in order to improve image segmentation~\cite{Chen2018} and music source separation~\cite{Liu2019}, as well as speech denoising~\cite{8461819}.  Previous work demonstrated the benefit of short-range dilated convolution for co-channel speaker identification~\cite{Wang2018}, but was unable to leverage dilated convolution effectively for the streaming of consecutive syllables, because vanishing gradients limited the number of dilated convolutional layers that could be effectively trained.  Better gradient propagation, and a much deeper stack of dilated convolutions, are possible by adding a residual connection every 3 convolutional layers in the dilated convolutional network; besides eliminating the vanishing gradient problem, residual connections help the network to maintain the focus on local time-frequency magnitude information while aggregating context.

\begin{figure}[h]
  \centering
  \includegraphics[width=\linewidth]{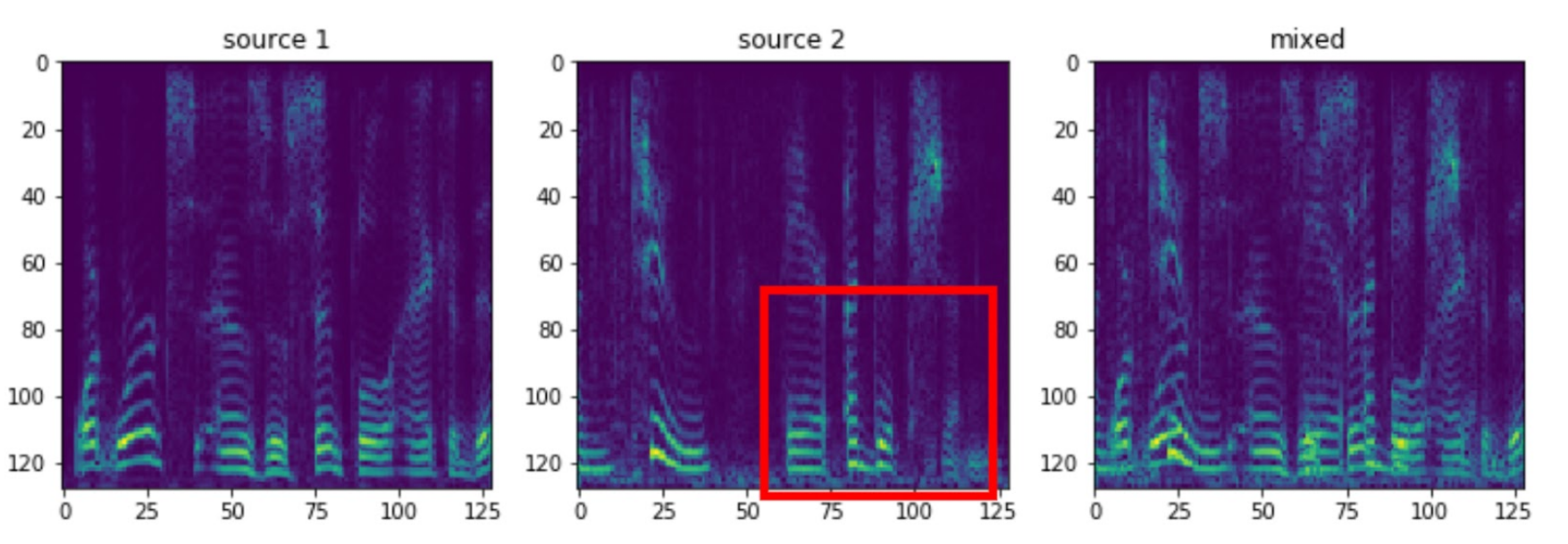}
  \caption{Example spectrograms of target source 1, target source 2, and the mixed signals.}
  \label{fig:2}
\end{figure}
\begin{figure}[h]
  \centering
  \includegraphics[width=0.7\linewidth]{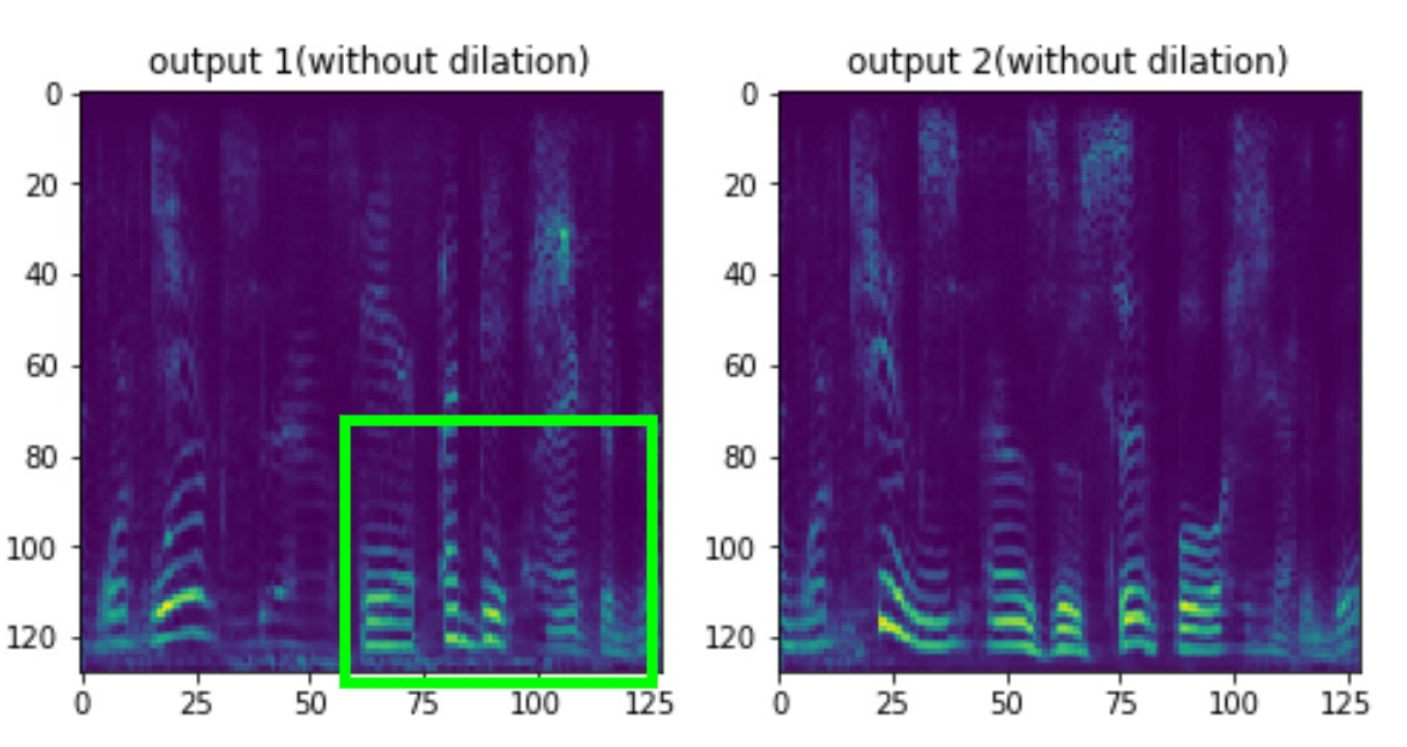}
  \caption{Example spectrograms: prediction without dilated convolution. It is obvious that left \& right halves of both output channels correspond to different sources.}
  \label{fig:3}
\end{figure}
\begin{figure}[h]
  \centering
  \includegraphics[width=0.7\linewidth]{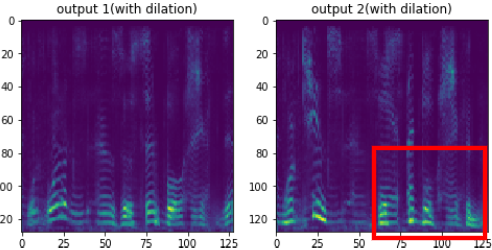}
  \caption{Example spectrograms: prediction with 5 blocks of dilated convolution, from the ablation study.}
  \label{fig:4}
\end{figure}

Figure \ref{fig:3} shows the result of extraction without using dilation with Figure \ref{fig:2} as input. It can be seen that the left and right parts of the extracted spectrogram come from mismatched sources. Figure \ref{fig:4} shows the result of extraction with 5 blocks of dilated convolution, and demonstrates that the mistake in Figure~\ref{fig:3} has been corrected, because of the increased receptive field provided by the dilated convolutions. Several architectures were tested; the best-performing system uses 3 dilated convolution blocks with $32\times\#\text{talker}$ channels, where  $\text{\#talker}$ is the number of talkers in each mixture. each with 6 dilated convolution layers, with the $n^{\textrm{th}}$ layer having dilation of $2^{n-1}$ pixels.


Residual attention and dilated convolution are pre-trained, prior to training the speaker identification  block, using permutation-invariant~\cite{Yu2017} mean squared error (MSE) of the reconstructed source spectrograms,
\begin{equation}
    MSE(\hat{S},S) = \min_{\substack{{\sigma\in \textrm{permutations}}\\(1:n_\textrm{sources})}}\sum_{i=1}^{n_{\textrm{sources}}} \Vert S_i-\hat{S}_{\sigma(i)}\Vert_2^2
\end{equation}
where $\hat{S}_c$ is the network output in channel $c$, and $S_i$ is the ground truth spectrogram of the $i^{\textrm{th}}$ source.

\section{Siamese speaker classification models}
\label{sec:classification}

Each output channel of the extractor model corresponds to the estimated spectrogram for a single source.  Each channel is classified using a Siamese (shared weights) speaker detection model~\cite{Chen2011}, in order to generate a prediction vector.  The output of the Siamese networks is a matrix of speaker identification probabilities, $\hat{y}_{i,c}=\Pr\left\{\mbox{channel}~c~\mbox{is speaker}~i\right\}$, with weights shared across the channel index, $c$.  Two different architectures were tested for the Siamese networks: syllable-level LSTM, and ResNet34.

\begin{figure}[h]
  \centering
  \includegraphics[width=0.65\linewidth,  keepaspectratio]{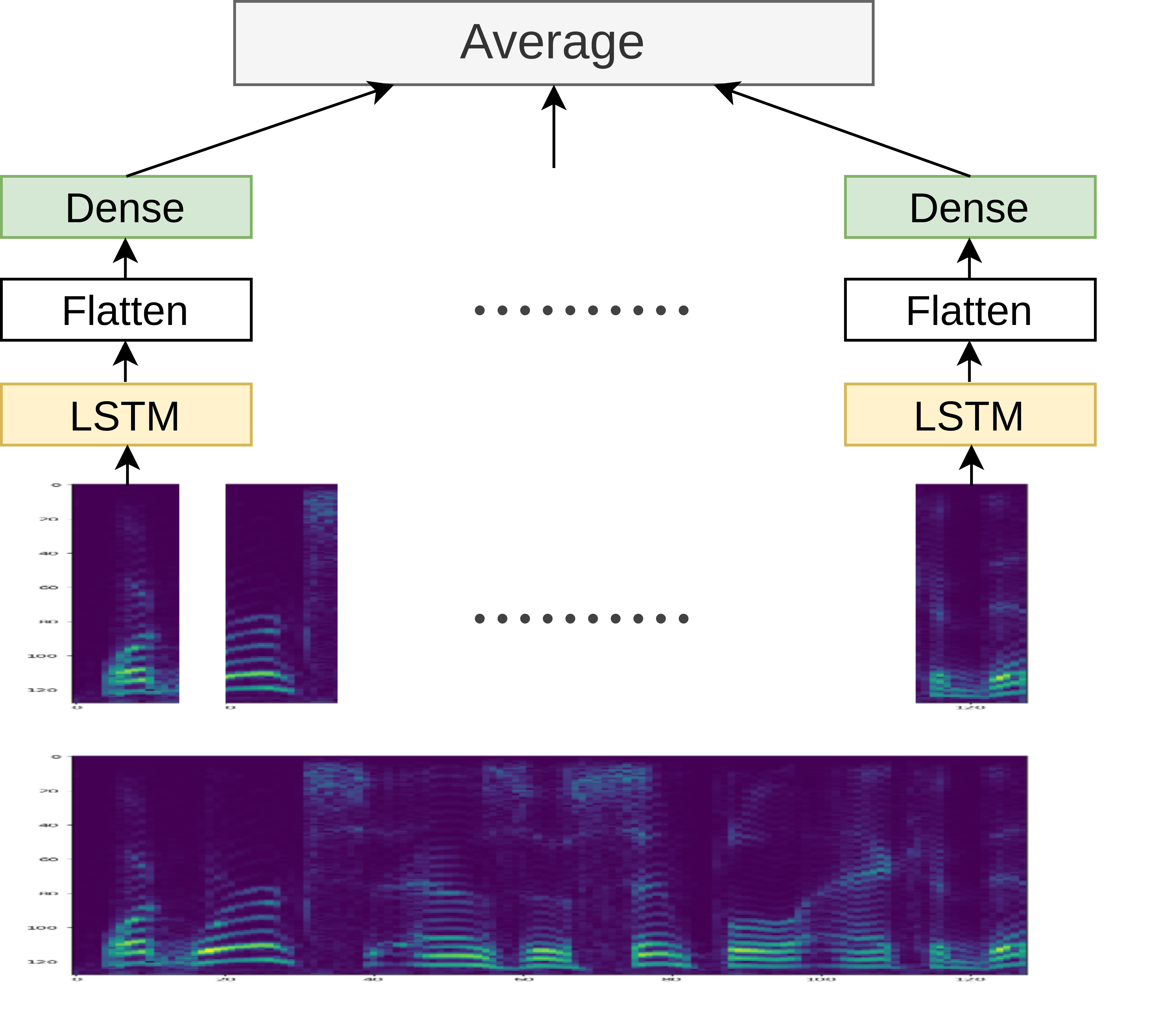}
  \caption{Syllable-level LSTM}
  \label{fig:5}
\end{figure}

The syllable-level LSTM is computed by chopping the extracted spectrogram into chunks of 312.5ms, which is about the length of a syllable, with a shift of 15.6ms between each chunk (Fig.~\ref{fig:5}). For each chunk, a 3-layer bi-LSTM with a hidden layer size of 200 computes features, which are concatenated into a single vector per chunk. Each chunk is classified using a 3-layer fully connected network (hidden layer size of 800, output size equal to the number of speakers), whose softmax outputs are averaged over the whole signal. By limiting window size to 312.5ms and averaging over the signal, we reduce the chance of extraction error affecting the prediction result by preventing it from changing the hidden state of the LSTM over a long range.

The ResNet34 classifier is a direct implementation of Resnet34~\cite{He2015}, with softmax activation at outputs. We found that for the global pooling layer, average pooling works significantly better than max pooling, probably due to the suppression of misassigned phonemes in the averaging operation.

In order to make the speaker identification network robust to output noise in the extracted spectrograms, the network is trained directly using extracted spectrograms as inputs. However, if we directly use the permutation decided by PIT~\cite{Yu2017} as input-target pairs, the accuracy of speaker classification is upper-bounded by the accuracy with which PIT is able to match extractor channels to reference sources. This is problematic especially in cases with more than 2 speakers, because PIT uses a minimum-MSE criterion to match output-reference pairs, but MSE does not directly measure speaker identification accuracy. Incorrect matching caused by PIT can result in the speaker classification network being trained on the wrong input-target pair, which will harm the accuracy of the system.

To address this issue, we propose channel-wise max pooling as a new way of aggregating output for such systems. To produce the final prediction vector for a mixed spectrogram, for each prediction index, we select its value from the Siamese network that yields the highest value. This is equivalent to stacking the Siamese network outputs in the channel dimension and performing channel-dimension max pooling. We then minimize the categorical cross entropy (CCE),
\begin{equation}
  CCE(\hat{y},y)=\sum_{i=1}^{n_{\textrm{speakers}}}-y_i\log\left(\max_{c=1}^{n_{\textrm{channels}}}\hat{y}_{i,c}\right)
  \label{eq:cce}
\end{equation}
between the predicted vector and ground truth, where $y_i=1$ when speaker $i$ is among the target speakers, and $n_{\textrm{speakers}}$ is the number of speakers known to the system. By letting the neural network vote for the match between ground truth speakers and predicted speakers, using the max-pooling framework shown in Eq.~(\ref{eq:cce}), we bypass the upper limit in the accuracy of PIT matching.

\section{Joint training of extractor and classifier with end-to-end back-propagation}
\label{sec:e2e}

\begin{figure}[h]
  \centering
  \includegraphics[width=0.65\linewidth]{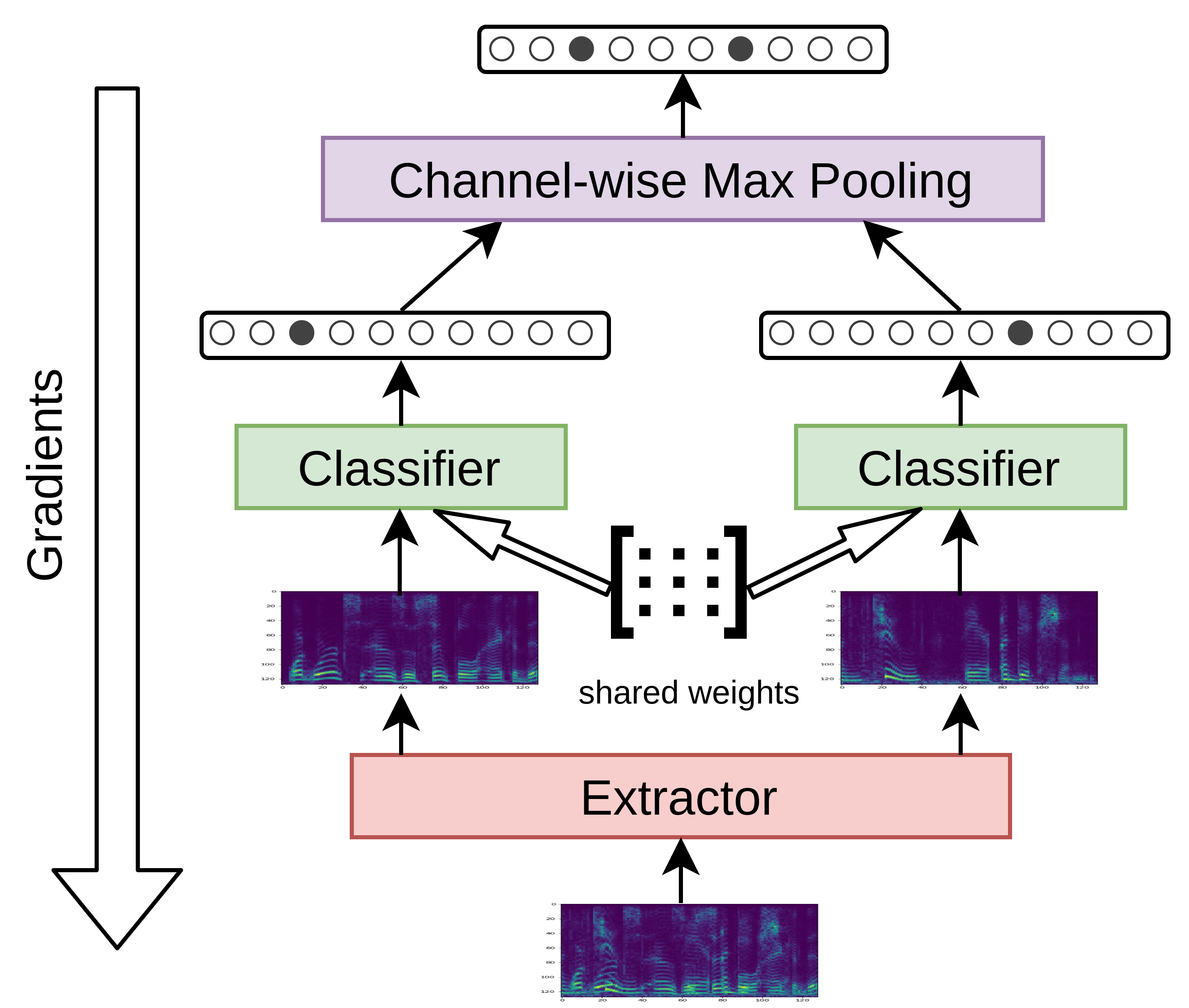}
  \caption{Joint training of system}
  \label{fig:6}
\end{figure}

Training the extractor model to minimize MSE is sensible if the training data contain clean, noise-free recordings of each target speaker.  Target speakers recorded from Broadcast News data, however, may be corrupted by music or other background noise signals, which are irrelevant to the task of speaker identification; an extractor trained in order to minimize MSE in the reconstruction of such sources will learn to waste parameters on signal components that are hard to estimate, and irrelevant to the task of speaker identification. To address this issue, we use joint training as shown in Fig.~\ref{fig:6} to further maximize the accuracy by pushing the network to place emphasis on features that help speaker classification. 

Since speaker classification loss is differentiable w.r.t. the separated spectrograms, its gradients can be back-propagated to the extractor model.  The extractor model is then trained using a two-part loss function, including CCE of the classifier, and scaled MSE of the extracted spectrograms:
\begin{equation}
  \mathcal{L}=\alpha\mbox{MSE}(\hat{S},S)+\mbox{CCE}(\hat{y},y)
  \label{eq:joint}
\end{equation}
We set $\alpha=20$ in 2 speaker case, and $\alpha=300$ in 3 speaker case, to level $\alpha\mbox{MSE}(\hat{S},S)$
and $\mbox{CCE}(\hat{y},y)$ to similar scales. 

\section{Experimental methods and results}
\label{sec:methods}

Training, development, and test data were extracted from Hub-4 broadcast news recordings~\cite{Graff2002}. We select top 20 speakers with the longest speech time, and for each person, extract all the non-overlapping 2s segments. All segments on average have  6.2\% silence time at -30dB threshold. We split those segments into 3 groups of 38424 training segments, 4803 validation segments, and 4886 test segments. During training/testing time, for each source mixture, we deterministically select the first source by iterating the dataset, select the other sources randomly from the remaining speakers, and add the signals to produce a mixture. 128-dimensional magnitude spectrum of each mixture audio is computed with window size of 32ms and hop size of 16ms.

Two versions of the proposed end-to-end system were tested. Both versions use the same extractor model (Sec.~\ref{sec:extractor}), but different classifiers (Sec.~\ref{sec:classification}): one of the classifiers is a syllable-level LSTM, one is ResNet34. The extractor is pre-trained using  MSE, and the classifier using CCE, then the extractor and classifier are pipelined, and trained using end-to-end back-propagation of the joint loss (Eq.~(\ref{eq:joint})).

Two baselines were implemented.  First, among previously published co-channel speaker identification systems, the one most comparable to ours is the system of Wang2018~\cite{Wang2018}: that system uses three layers of dilated convolution, but no residual connections or residual attention, and directly estimates speakers from mixtures without reconstructing sources.  Experimental tests reported in~\cite{Wang2018} demonstrated its success in separating laboratory recordings with predictable lexical content, but it has not been tested on Broadcast News recordings.  Second, we implemented a 2-stage modular baseline that we built ourselves with state-of-the-art networks. Conv-Tasnet~\cite{Luo2019} is used as extractor while SincNet is used as speaker classifier. Conv-Tasnet is first trained on the same training mixtures using PIT loss. Extracted signals are then paired with ground truth input sources using PIT in order to match each extracted signal with a target speaker, and SincNet~\cite{Ravanelli2018} is trained using the matched pairs. Training uses non-overlapping 0.2s chunks that cover the whole 2s signal as training inputs: SincNet is applied to each chunk, and its softmax outputs are averaged over time. Testing uses 0.2s chunks with 10ms shift, and time averaging of SincNet softmax outputs. Joint training is not used.

We also did an ablation study with the ResNet34 version of our pipeline. We replaced our residual attention blocks with dilated convolutional blocks, with the same number of channels, to see how much residual attention increases the accuracy.


\begin{table}
  \caption{Results; column $M/N$ lists \% of test data where at least $M$ talkers out of $N$-talker mixture are correctly identified.}
  \label{tab:results}
  \centering
  \begin{tabular}{|p{1.2cm}|c|c|c|c|c|c|}\hline
     \multicolumn{4}{|c|}{(\# spkr corr)/(\# in mixture)}&(s)&($10^9$)&($10^6$)\\
    Algorithm & 1/2 & 2/2 & 3/3&time&FLOPS&\#params \\\hline\hline
    Proposed, LSTM&99.9& 93.6 &77.7&41&41.1&12.4\\\hline
    Proposed, ResNet &99.9& 93.9 &81.2&47&158.6&20.1\\\hline
    TasNet + SincNet & 99.7 & 91.0 &74.1&52&23.8&23.0\\\hline
    Wang2018 & 95.2 & 36.3 & 12.5 & 2 & 0.15 & 1.36\\\hline
    Ablation & 99.8 & 92.1 &-&50&162.1 &17.4\\\hline
  \end{tabular}
\end{table}

Table~\ref{tab:results} lists experimental results.  The columns headed by $M/N$ list the percent of test data where at least $M$ speakers out of an $N$ speaker mixture are correctly identified.  The columns headed ``time,'' ``FLOPS,'' and ``\#params'' list the computation time, number of floating point operations, and number of parameters instantiated for the recognition of 4886 two-speaker mixtures with a batch size of 32 on $4\times$GTX 1080 Ti.  The Wang2018 baseline system has far fewer parameters than other systems, and is unable to recognize both speakers in a mixture of Broadcast News speech.  For the 2-speaker case, the modular baseline system, TasNet+SincNet, performs better, with correct recognition of both speakers in 91.0\% of test data.  The ablation study performs slightly better (92.1\%) than TasNet+SincNet, and does not approach the accuracies of the complete system (93.6\% and 93.9\% with syllable-level LSTM and ResNet34, respectively), suggesting that dilated convolution alone is insufficient for this task: the residual attention modules are critical for achieving high accuracy in this real-world scenario.  For the 3-speaker case, the proposed systems(77.7\% with syllable-level LSTM, 81.2\% with ResNet34) also outperformed the modular baseline(74.1\%) We also found that joint training brings the reconstruction MSE loss down, from 0.19 to 0.16 in the 2-speaker case, and from 0.32 to 0.27 in the 3-speaker case, possibly because extractor learns speaker-dependent features from classification loss gradients.

\section{Conclusions}
\label{sec:conclusion}

This paper presents a co-channel speaker identification system using residual attention, deep dilated convolution, and end-to-end optimization. The proposed system outperforms pipelines built with state-of-the-art neural networks: jointly optimizing the model, and using residual attention for the extractor, lead to better performance than cascading state-of-the-art source separation and speaker recognition systems.
All three systems achieve near-perfect accuracy in identifying at least one speaker out of two, therefore, it can be inferred that most errors in multi-speaker classification come from imperfect source reconstruction.  The proposed system avoids source reconstruction errors using a spectrogram masking method optimized by the back-propagation of speaker identification errors, and by using an 18-layer dilated convolution  with residual connections in order to enforce correct streaming of the component speech signals.
Compared to the TasNet+SincNet baseline, the proposed model has fewer channels and smaller kernels, therefore a smaller number of parameters and faster runtime possibly due to better parallelism.


\section{Acknowledgements}
This work was funded in part by NIH 1 R34 DA050256.

\bibliographystyle{IEEEtran}

\bibliography{mybib}

\end{document}